\documentclass[12pt,amssymb,amsfonts]{article} 
\usepackage{latexsym}

\newcommand {\be}{\begin{equation}}
\newcommand {\ee}{\end{equation}}
\newcommand{\bea}{\begin{eqnarray}}
\newcommand{\eea}{\end{eqnarray}}
\newcommand{\ba}{\begin{array}}
\newcommand{\ea}{\end{array}}
\newcommand{\beq}{\begin{eqnarray*}}
\newcommand{\eeq}{\end{eqnarray*}}

\newcommand{\ds}{\displaystyle}

\newcommand{\A}{{\cal A}}
\newcommand{\Y}{{\cal Y}}

\newcommand{\U}{{\bf U}}
\renewcommand{\k}{{\bf k}}
\newcommand{\x}{{\bf x}}
\newcommand{\e}{{\bf e}}
\renewcommand{\H}{{\bf H}}
\newcommand{\E}{{\bf E}}
\newcommand{\V}{{\cal V}}

\begin{document}

\begin{center}
{\Large\bf  Quotient Equations and Integrals of Motion for
Vector Massless Field}

\bigskip
S.S. Moskaliuk

\medskip
Bogolyubov Institute for Theoretical Physics\\
Metrolohichna Str., 14-b, Kyiv-143, Ukraine, UA-03143\\
e-mail: mss@bitp.kiev.ua

\end{center}

\bigskip
\centerline{Abstract}

\medskip
In this article a group-theoretical aspect of the method of dimensional
reduction is presented. Then, on the base of symmetry analysis of an
anisotropic space geometrical description of dimensional reduction of
equations for vector massless field is given. Formula for calculating
components of the energy-momentum tensor from the variables of the
field factor-equations is derived.

\newpage

The covariant equations for massless vector field in the absence of
sources are of the form

\be
\nabla_\mu F^{\mu\nu}=0, \qquad \nabla_\mu(*F)^{\mu\nu}=0,
\ee
where $F^{\mu\nu}$ is tensor of vector field, $(*F)^{\mu\nu}$ is
conjugate magnitude defined by the equation

\beq
(*F)^{\alpha\beta} ={1\over2\sqrt{-g}}
[\alpha\beta\gamma\eta]F_{\gamma\eta};
\eeq
$[\alpha\beta\gamma\eta]$ is completely antisymmetric tensor with
[0123]=1.

Because metric tensor ($g_{\mu\nu}$) does not depend on space
coordinates, the theory under consideration is invariant under space
translations, and consequently, the space transformation of
Fourier-components of tensor is rendered possible:

\bea
F^{\mu\nu}(t,\x) &\!\!\!=\!\!\!&
\int d^3k\exp(i, \k, \x) f^{\mu\nu}(t, \k),\nonumber \\
(*F)^{\mu\nu}(t, \x) &\!\!\!=\!\!\!&
\int d^3k\exp(i\k\x)(*f)^{\mu\nu}(t,\k).
\eea
Substituting (2) in (1), one comes to the equation for
$f^{\mu\nu}(t, \k)$:

\bea
&& -ik_jf^{0j} =0,\nonumber \\
&& -\left[ {1\over\sqrt{-g}}
\partial_0\left( \sqrt{-g}\, f^{j0}\right)
+ik_jf^{ji}\right]=0.
\eea
Because $f^{ji}=-[jri]\A_r^2(*f)^{r0}/\sqrt{-g}$, the previous equation
takes the form

\be
{d\over dt}\left(\sqrt{-g} f^{0j}\right) =-i[jml]
k_l\A_m^2(t)(*f)^{m0}.
\ee
Similarly, for the conjugate magnitude we have

\bea
&& k_j(*f)^{j0}=0, \nonumber \\
&& {d\over dt} \left(\sqrt{-g}(*f)^{j0}\right) =i[jml]
k_l\A_m^2(t)f^{0m}.
\eea

To simplify the set of equations (3)--(5), we introduce the new
variables

\be
S_j^\pm =\sqrt{-g}[f^{0j}\pm i(+f)^{j0}].
\ee
This enables us to separate the equations (3)--(5) for $S_j^+$ and
$S_j^-$:

\be
k_jS_j^\pm=0, \qquad {d\over dt}S_j^\pm =\mp {[jml]\over\sqrt{-g}} \,
k_l\A_m^2 S_m^\pm.
\ee
The further simplification consists of the introduction spherical
coordinates in momentum space according to the relation

\be
\left(k_1,k_2, k_3\right) =k\left(\sin(\delta)\cos(\xi),
\sin(\delta)\sin(\xi), \cos(\delta)\right).
\ee
The equation (7) can be automatically satisfied going from (6) to the
magnitudes

\bea
S_\delta^\pm  &\!\!\!=\!\!\!&
\cos(\delta)\cos(\xi)S_1^\pm +\cos(\delta)\sin(\xi)S_2^\pm
-\sin(\delta)S_3^\pm, \nonumber \\
S_\xi^\pm  &\!\!\!=\!\!\!&
-\sin(\xi) S_1^\pm +\cos(\xi) S_2^\pm.
\eea
>From the equation (7), rewritten as
\beq
\sin(\delta)\left( \cos(\xi) S_1^\pm +\sin(\xi) S_2^\pm\right)
+\cos(\delta) S_3^\pm=0,
\eeq
and definition of $S_\delta^\pm$ it is possible to derive the relations

\beq
S_\delta^\pm={1\over \cos(\delta)} \left( \cos(\xi) S_1^\pm
+\sin(\xi)S_2^\pm\right), \quad S_3^\pm =-S_\delta^\pm\sin(\delta).
\eeq
Considering the previous relations and the definition $S_\xi^\pm$ as a
set of equations, we obtain for $S_1^\pm$ and $S_2^\pm$ the following
coupling relations

\bea
S_1^\pm  &\!\!\!=\!\!\!&
S_\delta^\pm\cos(\xi)\cos(\delta) -S_\xi^+\sin\xi,\nonumber \\
S_2^\pm &\!\!\!=\!\!\!&
S_\delta^\pm\sin(\xi)\cos(\delta) +S_\xi^\pm\cos(\xi), \nonumber \\
S_3^\pm &\!\!\!=\!\!\!&  -S_\delta^\pm\sin(\delta).
\eea
To derive the equations for $S_\delta^\pm$ and $S_\xi^\pm$, let us
differentiate the amgnitude (9) by $t$:

\bea
\dot{S}_\xi^\pm  &\!\!\!=\!\!\!& \cos(\delta)\cos(\xi)
\dot{S}_1^\pm +\cos(\delta)\sin(\xi)\dot{S}_2^\pm -\sin(\delta) 
\dot{S}_3^\pm, \nonumber \\ 
\dot{S}_\xi^\pm   &\!\!\!=\!\!\!& 
-\sin(\xi) \dot{S}_1^\pm +\cos(\xi) \dot{S}_2^\pm .  \eea

Further we use the equations (7)

\beq
\pm\dot{S}_1^\pm
 &\!\!\!=\!\!\!&
-{k\over\sqrt{-g}}\left(\cos(\delta) \A_2^2
{S}_2^\pm -\sin(\delta)\sin(\xi)\A_3^2
{S}_3^\pm\right),  \\
\pm\dot{S}_2^\pm
 &\!\!\!=\!\!\!&
+{k\over\sqrt{-g}} \left( \cos(\delta)\A_1^2
{S}_1^\pm -\sin(\delta)\cos(\xi)\A_3^2
{S}_3^\pm\right),  \\
\pm \dot{S}_3^\pm
 &\!\!\!=\!\!\!&  {k\over\sqrt{-g}}
\left( \sin(\delta)\sin(\xi)\A_1^2
{S}_1^\pm -\sin(\delta)\cos(\xi)\A_2^2
{S}_2^\pm\right),
\eeq
and then (10) with the result that for
${S}_\delta^\pm$ and ${S}_\xi^\pm$
we obtain the equations

\bea
\pm \dot{S}_\delta^\pm  &\!\!\!=\!\!\!&
-kaS_\delta^\pm -kb{S}_\xi^\pm,\nonumber \\
\pm \dot{S}_\xi^\pm  &\!\!\!=\!\!\!&
+kc {S}_\delta^\pm +ka{S}_\xi^\pm.
\eea
The parameters $a$, $b$ and $c$ can be presented as follows
\bea
a &\!\!\!=\!\!\!&  {\cos(\delta)\cos(\xi)\sin(\xi)\over
\sqrt{-g}} \left( \A_2^2(t) -\A_1^2(t)\right),\nonumber \\
b &\!\!\!=\!\!\!&  {1\over\sqrt{-g}} \left(\A_2^2(t) \cos^2(\xi)
+\A_1^2(t)\sin^2(\xi)\right),\nonumber \\
c &\!\!\!=\!\!\!& {1\over\sqrt{-g}} \left(\A_1^2(t)
\cos^2(\delta)\cos^2(\xi) +\A_2^2(t) \cos^2(\delta) \sin^2(\xi)
+\A_3^2(t)\sin^2(\delta) \right).
\eea
To continue the analysis, it is necessary to discuss the dependence of
basis vectors of space on time. The metric $g_{\mu\nu}$ defines the
natural covariant "unit" vector $\e_\mu$ with covariant components
$e_\mu^\alpha=\delta_\mu^\alpha$ and the natural contravariant vector
$\e^\mu$ with contravariant components $e_\alpha^\mu=\delta_\alpha^\mu$
with $(\e_\mu \e_\nu)=g_{\mu\nu}$.
The absence of the off-diagonal terms from metric means that the basis
consisting of the vectors $\e_\mu$ is orthogonal, but the length of
space basis vectors varies with time according to the relation

\beq
\left( \e_i \e_i\right)=\A_i^2(t).
\eeq
Hence it follows that the components of various tensors calculated in
this basis do not carry complete information on the corresponding
fields. The difficulty mentioned above can be easily avoided by
introduction of tetrade basis vectors according to the relations
$\e_{(0)}=\e_0$, $\e_{(i)}=\e_i/\A_i(t)$. It turns out that, doing so, 
we obtain $(\e_{(\mu)} \e_{(\nu)})=\eta_{\mu\nu}$.

In the case under consideration, because $g_{\mu\nu}$ does not depend on
space coordinates, the tetrade basis can be introduced for all points
of space at any given instant of time.

The vectors of electric and magnetic field strength can be written in
tetrade basis as follows

\bea
\E(t, \x) &\!\!\!=\!\!\!& \int d^3 k\, \exp(i\k\x)\sum_{i=1}^3\A_i(t)
f^{0i}(t, \k)\e_{(i)},\nonumber \\
\H(t, \x) &\!\!\!=\!\!\!&  \int d^3 k\, \exp(i\k\x)\sum_{i=1}^3\A_i(t)
(*f)^{0i}(t, \k) \e_{(i)}.
\eea
The spectral components $\E$ and $\H$, calculated in tetrade basis, are
orthogonal to the direction, varying with time, which is determined by
vector $k_i=k_i/\A_i(t)$. In this connection it is convenient to
introduce the unit vectors varying with time

\bea
\e_k &\!\!\!=\!\!\!& \sin(\theta)\,
\cos(\varphi)\e_{(1)}+\sin(\theta)\sin(\varphi)\e_{(2)} 
+\cos(\theta)\e_{(3)}, \nonumber \\ 
\e_\theta &\!\!\!=\!\!\!& 
\cos(\theta)\, \cos(\varphi)\e_{(1)} +\cos(\theta)\, 
\sin(\varphi)\e_{(2)} -\sin(\theta)\e_{(3)},\nonumber \\ 
\e_\varphi 
&\!\!\!=\!\!\!& -\sin(\varphi)\e_{(1)}+\cos(\varphi)\e_{(2)}, 
\eea 
where the angles $\theta$ and $\varphi$ are defined by the relation

\bea
&& \left(\sin(\theta)\cos(\varphi),
\sin(\theta)\sin(\varphi),\cos(\theta)\right)=\nonumber \\
 &\!\!\!=\!\!\!& \mu^{-1} \left({\sin(\delta)\cos(\xi)\over\A_1}, \,
{\sin(\delta)\sin(\xi)\over\A_2}, \, {\cos(\delta)\over\A_3} \right),\\
\mu &\!\!\!=\!\!\!& \left( {\sin^2(\delta)\cos^2(\xi)\over\A_1^2}
+{\sin^2(\delta)\sin^2(\xi)\over\A_2^2}
+{\cos^2(\delta)\over\A_3^2}\right)^{1/2}.\nonumber
\eea
Let us transform the expression (14), using the functions
$S_\delta^\pm$, $S_\xi^\pm$, introduced earlier. We shall do this for
magnetic field $\H$. Using (16), let us present $\H$ in the form

\beq
\H(t, \x) &\!\!\!=\!\!\!& \int d^3 k\exp(i\k\x) \sum_{l=1}^3 \,
{1\over2i\sqrt{-g}} \left(S_l^+ -S_l^-\right) \A_l(t) \e_{(l)}=\\
 &\!\!\!=\!\!\!& {1\over 2i\sqrt{-g}}\, \sum_{r=\pm1} \, \int d^3
 k\exp(i\k\x) r\sum_{l=1}^3 S_l^r\A_l(t)\e_{(l)}.
\eeq
In the last relation we went from the notation $S_l^\pm$ to the
notation $S_l^r$, where $r=\pm1$. Let us consider the internal sum. The
change $S_l^\pm\rightarrow S_\delta^r$ and $S_\xi^r$ according to (10)
brings it to the form

\beq
&& \sum_{l=1}^3 S_l^r\A_l(t)\e_{(l)} = S_\xi^r
\left[-\sin(\xi)\A_1\e_{(1)} +\cos(\xi)\A_2\e_{(2)}\right]  +\\
 &\!\!\!+\!\!\!& S_\delta^r\left[ \cos(\xi)\cos(\delta)\A_1 \e_{(1)}
+\sin(\xi)\cos(\delta)\A_2 \e_{(2)} -\sin(\delta)\A_3 \e_{(3)} \right].
\eeq
Let us go from the angles $\xi$, $\delta$ to the angles $\varphi$,
$\theta$. To this end we shall use the following transition formulas,
which can be derived from (16):

\bea
\sin(\varphi) &\!\!\!=\!\!\!&  {\A_1\over (-g)^{1/4} b^{1/2}}\,
\sin(\xi), \qquad \cos(\varphi) ={\A_2\over (-g)^{1/4} b^{1/2}}\,
\cos(\xi), \nonumber \\
\sin(\theta) &\!\!\!=\!\!\!&  {\A_3 b^{1/2}\over \mu(-g)^{1/4}}\,
\sin(\delta), \qquad \cos(\theta) ={1\over \mu^{-1}\A_3}\,
\cos(\delta).
\eea
As a result, taking into account the definition $\e_\varphi$, we obtain

\beq
\sum_{l=1}^3 S_l^r\A_l(t) \e_{(l)} 
 &\!\!\!=\!\!\!& {(-g)^{1/4}\over b^{1/2}} \left[
S_\delta^r \mu\left( {\A_1\A_3\over\A_2} b\cos(\varphi) \cos(\theta)
\e_{(1)}+ \right. \right. \\
 &\!\!\!+\!\!\!& \left.\left.
{\A_2\A_3\over\A_1} b\sin(\varphi)\cos(\theta) \e_{(2)}
-\sin(\theta)\e_{(3)}\right) +S_\xi^r b \e_{(\varphi)}\right].
\eeq
It is easy to verify that the following eqution is satisfied

\beq
&& \left( {\A_1\A_3\over\A_2} b-1\right) \cos(\varphi)\cos(\theta)
\e_{(1)} +\\
 &\!\!\!+\!\!\!& \left({\A_2\A_3\over\A_1} b-1\right)
\sin(\varphi)\cos(\theta)\e_{(2)}= {a\over\mu}\,  \e_\varphi.
\eeq
This can be done, using the relation
\beq
\left( {\sqrt{-g}\, b\over\A_1^2} -1\right) ={\A_1^2-\A_2^2\over
\A_2^2} \sin^2(\xi), \qquad \left( {\sqrt{-g}\, b\over\A_2^2} -1\right)
={\A_2^2-\A_1^2\over\A_1^2} \cos^2(\xi)
\eeq
and the connection formulas (17).

Then the next result is
\beq
\sum_{l=1}^3 S_l^r\A_l(t)\e_{(l)}  &\!\!\!=\!\!\!&  
{(-g)^{1/4}\over
b^{1/2}} \left[ S_\delta^r \left(\mu \e_\theta +a \e_\varphi\right) 
+S_\xi^r b \e_\varphi\right] =\\ 
&\!\!\!=\!\!\!& {(-g)^{1/4}\over 
 b^{1/2}} \left[ S_\delta^r \mu \e_\theta -{r\over k} \dot{S}_\delta^r 
\e_\varphi\right].  
\eeq 
The magnetic field strentgth can be written 
now as follows:

\be
\H={1\over 2i(-g)^{1/4}} \, \sum_{r=\pm1} \int d^3 k\exp(i\k\x) 
rb^{-1/2} \left[ S_\delta^r \mu \e_\theta -{r\over k} \dot{S}_\delta^r 
\e_\varphi\right].
\ee
For the vector of elctric field strength the calculations are carried
out in analogous way and give the following result:
\be
\E={4\over 2i(-g)^{1/4}} \, \sum_{r=\pm1} \int d^3 k\exp(i\k\x) rb^{-1/2}
\left[ S_\delta^r \mu \e_\theta -{r\over k} \dot{S}_\delta^r
\e_\varphi\right].
\ee

The formulas (18) and (19) bring out that the electric and magnetic
fields are completely described by the function $S_\delta^r$. That is
why it is convenient to go from (12) to the equation of the second
order for $S_\delta^r$. To this end let us differentiate the first
equation (12) by $t$, substitute the second equation and eliminate the
rest terms, using again the first equation. As a result, we obtain

\beq
{d\over dt}\dot{S}_\delta^r ={\dot{b}\over b} \dot{S}_\delta^r
+S_\delta^r \left(-kr\dot{a} +k^2(a^2-bc)+{\dot{b}\over b} ka\right).
\eeq
Using the fact that $a^2-bc=-\mu^2$, we come to
\be
\ddot{S}_\delta^r-{\dot{b}\over b} \dot{S}_\delta^r+
\left(k^2\mu^2+k\Lambda^r\right) S_\delta^r=0,
\ee
with $\Lambda^r=r(\dot{a}-{\dot{b}\over b}a)$.

Let us find out the restrictions on the functions $S_\delta^r$,
following from the real valuedness of the fields $\E$ and $\H$. It stems
from the conditions

\beq
F^{\alpha\beta}(t, \x) = 
\stackrel{\!\!\!\!\!\!*}{F^{\alpha\beta}}(t, \x), \qquad 
(*F)^{\alpha\beta}(t, \x) =(*\stackrel{*}{F})^{\alpha\beta}(t, \x) 
\eeq 
immediately the requirement that 
~~$f^{\alpha\beta}(t,\k)=\stackrel{\!\!\!\!\!\!*}{f^{\alpha\beta}}(t,\k)$,
~~~~~$(*f)^{\alpha\beta}(t,\k)=\\
=(*\stackrel{*}{f})^{\alpha\beta}(t,-\k)$.
Then it follows from definition (6) straightforwardly that
\be
S_j^+(t,\k)=
\stackrel{\!\!\!\!\!\!*}{S_j^-}(t,-\k), \qquad S_j^r(t,\k)=
\stackrel{\!\!\!\!\!\!*}{S_j^{-r}}(t,-\k).
\ee
Because the reflection in the momentum space corresponds to the change

\beq
&&\delta'=\pi-\delta, \quad \xi'=\pi+\xi, \quad
\cos(\delta')=-\cos(\delta), \quad \sin(\delta')=\sin(\delta),\\
&& \cos(\xi')=-\cos(\xi), \quad \sin(\xi')=-\sin(\xi),
\eeq
we obtain that
\beq
S_\delta^r(t,\k)=\stackrel{\!\!\!\!\!\!*}{S_\delta^{-r}}(t,-\k)
\eeq
(vectors $\e_\theta$ and $\e_\varphi$ under the change $\k$ for $-\k$
transform themselves in the same way).

Let us present $S_\delta^r$ as

\be
S_\delta^r=C_1^r(\k)\Y^r(t,\k)+C_2^r(\k)
\stackrel{\!\!\!\!\!\!*}{\Y^r}(t,\k).
\ee

This representation corresponds to the fact that $S_\delta^r$ satisfies
the equation of the second order (20).

The magnitudes $C_{1,2}^r(\k)$ are coefficients which depend on momentum
$\k$, and the functions $\Y^r(t,\k)$ carry the information about the
dependence on time $t$.

Let us find out the properties of $C_{1,2}^r(\k)$ and $\Y^r(t,\k)$.
First of all it is necessary to establish the connection between
$C_1^r$ and $C_2^r$. To this end we shall consider the isotropic case
$\A_1=\A_2=$ $=\A_3=R$. It is easy to verify that $a=0$, $b=1/R$,
$\mu=1/R$. Substituting these values in (20), we bring this equation to
the form
\beq
\ddot{S}_\delta^r+{\dot{R}\over R} \dot{S}_\delta^r +{k^2\over R^2}
S_\delta^r =0.
\eeq
It is easy to check up that the general solution of the latter is of
the form $S_\delta^r=C_1^r(\k)\exp(-ik\eta) +C_2^r(\k)\exp(ik\eta)$ with
$\eta=\int dt/R(t)$.

Then the equation (21) can be rewritten as

\beq
\stackrel{\!\!\!\!\!\!*}{S^+_\delta}(-\k) &\!\!\!=\!\!\!&
\stackrel{\!\!\!\!\!\!*}{C^+_1}(-\k)\exp(ik\eta)
+\stackrel{\!\!\!\!\!\!*}{C^+_2}(-\k)\exp(ik\eta),\\
\stackrel{\!\!\!\!\!\!*}{S^-_\delta}(\k) &\!\!\!=\!\!\!&
C^-_1(\k)\exp(-ik\eta) +C_2^-(\k)\exp(ik\eta),
\eeq
hence,
$\stackrel{\!\!\!\!\!\!*}{C_1^+}(-\k)=C_2^-(\k)$,
$\stackrel{\!\!\!\!\!\!*}{C_2^+}(-\k)=C_1^-(\k)$.

In general case of anisotropic metric the equation (21) is of the form

\beq
&&
\stackrel{\!\!\!\!\!\!*}{C_1^+}(-\k)\stackrel{\!\!\!\!\!\!*}{\Y^+}(-\k)+
\stackrel{\!\!\!\!\!\!*}{C_2^+}(-\k)\Y^+(-\k)=\\
 &\!\!\!=\!\!\!& C_1^-(\k)\Y^-(\k)+ C_2^-(\k)
\stackrel{\!\!\!\!\!\!*}{\Y^-}(\k),
\eeq
or
\beq
C_2^-(\k)
\stackrel{\!\!\!\!\!\!*}{\Y^+}(-\k)+ C_1^-(\k)\Y^+(-\k) = 
C_1^-(\k)\Y^-(\k) +C_2^-(\k)\stackrel{\!\!\!\!\!\!*}{\Y^-}(\k).  
\eeq The 
last equation can be easily satisfied by putting $\Y^+(-\k)=\Y^-(\k)$. 
In the isotropic case $\Y^\pm(\k)=C_0\exp(ik\eta)$, i.e. the last 
equation is satisfied trivially.

Let us substitute (22) in (19). It is easy to verify that

\beq
\sum_{r=\pm1} \, \left[ S_\delta^r \mu \e_\theta-{r\over
k} \dot{S}_\delta^r \e_\varphi\right] =\sum_{r=\pm1}\, \left[C_1^r(\k)
\U_0^r(t,\k) +C_2^r(\k)
\stackrel{\!\!\!\!\!\!*}{\U_0^r}(t,\k)\right]
\eeq
with
\beq
\U_0^r(t,\k) =\mu\Y^r \e_\theta-{r\over k}\dot{\Y}^r \e_\varphi.
\eeq
Then $\E$ equires the form

\beq
\E &\!\!\!=\!\!\!&  {1\over 2(-g)^{1/4}}\, \sum_{r=\pm1} \int d^3
k\exp(i\k\x) b^{-1/2}\left[ C_1^r(\k)\U_0^r(t,\k)+ \right.\\
 &\!\!\!+\!\!\!& \left. C_2^r(\k)
\stackrel{\!\!\!\!\!\!*}{\U_0^r}(t,\k) \right].
\eeq
The expression for $\E$ contains two integrals. Let us change in the
second integral the integration variable $\k$ for $-\k$ and consider the
relation

\beq
&&
\sum_{r=\pm1}\, C_2^r(-\k)\U_0^r(t,-\k)=\\
 &\!\!\!=\!\!\!& \sum_{r=\pm1}\, C_2^r(-\k)
\left( \mu(-\k)-\Y^r(t,-\k) \e_\theta(-\k) -{r\over k}\dot{\Y}^r(t,-\k)
\e_\varphi(-\k)\right)=\\
 &\!\!\!=\!\!\!& \sum_{r=\pm1} \, C_1^{-r}(\k)
\left(\mu(\k)
\stackrel{\!\!\!\!\!\!*}{\Y^{-r}}(t,\k) \e_\theta(\k)+ {r\over k}
\stackrel{\!\!\!\!\!\!*}{\dot{\Y}^{-r}}(t,\k)
\e_\varphi(\k)\right)=\\
 &\!\!\!=\!\!\!& \sum_{r=\pm1}\,
\stackrel{\!\!\!\!\!\!*}{C_1^r}(\k)
\stackrel{\!\!\!\!\!\!*}{\U_0^r}(t,\k).
\eeq
After doing so, the expression for $\E$ acquires the form

\bea
\E &\!\!\!=\!\!\!&  {1\over 2(-g)^{1/4}}\, \sum_{r=\pm1} \int d^3k b^{-1/2}
\left[ C_1^r(\k) \U_0^r(t,\k)\exp(i\k\x)+ \right. \nonumber \\
 &\!\!\!+\!\!\!& \left.
\stackrel{\!\!\!\!\!\!*}{C_1^r}(\k)
\stackrel{\!\!\!\!\!\!*}{\U_0^r}(t,\k)\exp(-i\k\x)\right.\, .
\eea

Let us present $\E$ as expansion in respect to complete system of
eigenfunctions of the corresponding equation with amplitudes $C_1^r(\k)$
(the index 1 shall be further omitted).

The vectors $\U_0^r(t,\k)$, involved in the expansion (23), in the case
of anisotropic space are analogues of the vectors determining a certain
state of circular polarization (in quantum language -- determining the
states of particles with a certain projection of spin on the direction
of movement).

The expansion for magnetic field $\H$ similar to (23) is of the form

\bea
\H &\!\!\!=\!\!\!& {1\over 2i(-g)^{1/4}}\, \sum_{r=\pm1}\, 
\int d^3k rb^{-1/2} \left[ C^r(\k)\U_0^r(t,\k) \exp(i\k\x)+\right. 
\nonumber  \\ 
&\!\!\!+\!\!\!& \left.  \stackrel{\!\!\!\!\!\!*}{C^r}(\k) 
\stackrel{\!\!\!\!\!\!*}{\U_0^r}(t,\k) \exp(-i\k\x) \right].
\eea
Further we shall modify (23) and (24), introducing in the definition of
the vectors $\U_0^r(t,\k)$ all functions of time:

\beq
\U^r(t,\k) ={1\over(-g)^{1/4} b^{1/2}} \left( \mu\Y^r \e_\theta -{r\over
k} \dot{\Y} \e_\varphi\right).
\eeq
The ultimate expansion for electric and magnetic fields (EMF) is as
follows:

\bea
\E &\!\!\!=\!\!\!& {1\over 2} \sum_{r=\pm1}\, \int d^3 k \left[
C^r(\k)\U^r(t,\k)\exp(i\k\x) + \right.\nonumber \\
 &\!\!\!+\!\!\!& \left.
\stackrel{\!\!\!\!\!\!*}{C^{+(r)}}(\k)
\stackrel{\!\!\!\!\!\!*}{\U^r}(t,\k)\exp(-i\k\x)\right],\nonumber 
\\[0.3cm] 
\H &\!\!\!=\!\!\!& {1\over2}\sum_{r=\pm1}\, \int d^3 k r \left[ 
C^r(\k)\U^r(t,\k) \exp(i\k\x)+ \right. \nonumber \\
 &\!\!\!+\!\!\!& \left.
\stackrel{\!\!\!\!\!\!*}{C^{+(r)}}(\k)
\stackrel{\!\!\!\!\!\!*}{\U^r}(t,\k)\exp(-i\k\x)\right].
\eea

The main characteristic of quantum effects of EMF in anisotropic space
are vacuum averages of the operators of symmterized energy-momentum
tensor (EMT) of EMF

\beq
T_{\alpha\mu}(t) &\!\!\!=\!\!\!&
\langle 0|N_t\hat{T}_{\alpha\mu}(t,x)|0\rangle,\\
N_t\hat{T}_{\alpha\mu} &\!\!\!=\!\!\!& \hat{T}_{\alpha\mu}-\langle
0_t|\hat{T}_{\alpha\mu}|0_t\rangle,\\
\hat{T}_{\alpha\mu} &\!\!\!=\!\!\!& -{g^{\nu\rho}\over2}
\left\{\hat{F}_{\mu\rho}, \hat{F}_{\alpha\nu}\right\}+
{1\over8} g_{\alpha\mu} \left\{ \hat{F}^{\beta\nu}, \hat{F}_{\beta\nu}
 \right\}, \\
\left\{\hat{A}, \hat{B}\right\}  &\!\!\!=\!\!\!&
 \hat{A}\hat{B}-\hat{B}\hat{A}.
 \eeq
In particular, $T_0^0(t)={1\over\V}\langle
0|N_t\hat{\H}(t)|0\rangle$. Similarly to the fact that the solution of
Maxwell equations can be presented as superposition of plane waves with
wave vector $\k$, EMT can be expanded as follows:

\be
T_\nu^\mu=\int d\xi d\delta \sin(\delta)\int d K_0(t,k,\delta,\xi)
\widetilde{T}_\nu^\mu(t,k,\delta,\xi),
\ee
where $K_0(t,k,\delta,\xi)=k\mu(t,\delta,\xi)$ is photon physical
frequency.

The nonzero spectral components of (26) are

\beq
\widetilde{T}_0^0={k^3\over\V}\sum_{r=1}^2  2S^r,
\eeq
\bea
\widetilde{T}_1^1  &\!\!\!=\!\!\!&
{k^3\over\V}\sum_{r=1}^2 \left[-\cos(2\varphi)X^r+\sin(2\varphi)Y^r
-{\sin^2(\theta)\over2} \left(2S^r+U^r\right)\right],\nonumber \\
\widetilde{T}_2^2  &\!\!\!=\!\!\!&
{k^3\over\V}\sum_{r=1}^2
\left[\cos(2\varphi)X^r-\sin(2\varphi)Y^r
-{\sin^2(\theta)\over2} \left(2S^r+U^r\right)\right],\nonumber \\
\widetilde{T}_3^3  &\!\!\!=\!\!\!&
{k^3\over\V}\sum_{r=1}^2 \left[-\cos(\theta)2S^r
+\sin^2(\theta)U^r
\right],\nonumber \\
\widetilde{T}^{12}  &\!\!\!=\!\!\!&
{k^3\over A_1A_2\V}\sum_{r=1}^2
\left[\sin(2\varphi)X^r+\cos(2\varphi)Y^r\right],\nonumber \\
\widetilde{T}^{13}  &\!\!\!=\!\!\!&
{k^3\over A_1A_3\V}\sum_{r=1}^2
\left[\cos(\varphi){\sin(2\theta)\over2}
\left(2S^r+U^r\right)+\sin(\varphi){\rm tg}(\theta)Y^r\right],\nonumber
\\ \widetilde{T}^{23}  &\!\!\!=\!\!\!& {k^3\over A_2A_3\V}\sum_{r=1}^2
\left[\sin(\varphi){\sin(2\theta)\over2}
\left(2S^r+U^r\right)-\cos(\varphi){\rm tg}(\theta)Y^r\right].
\eea
The components $T^{0i}$ vanish due to spaceous homogeneity of metric.
The trace $\widetilde{T}_\mu^\mu$ of (27) vanishes, because the field
is massless. The condition of conservativeness $\nabla_\mu T^{\mu0}=0$
is satisfied, which in the chosen metric is substantial only for
diagonal components of EMT.

In contrast to the case of isotropic space, where EMT is diagonal for
all types of fields, EMT in anisotropic space is nondiagonal in respect
to space indices. This is related with nonorthogonality of polarization
vectors $\U^{(_1)}(t,\k)$ and $\U^{(-1)}(t,\k)$:

\beq
&&
\left(\U^{(+1)}(t,\k), \stackrel{\!\!\!\!\!\!*}{\U^{(-1)}}(t,\k)\right)=
b^{-1}(t) \left(-g(t)\right)^{-1/2} \left(\mu^2(t,\k) \Y^{+1}(t,\k)
\right. \times \\
&&\left. \times \stackrel{\!\!\!\!\!\!*}{\Y^{-1}}(t,\k)-{1\over k^2}
\dot{\Y}^{+1}(t,\k) 
\stackrel{\!\!\!\!\!\!*}{\dot{\Y}^{-1}}(t,\k)\right).
\eeq
(At $t=t_0$, according to initial conditions, $(\U^{(+1)},
\stackrel{\!\!\!\!\!\!*}{\U^{(-1)}})=0$.) For  this reason it is 
impossible to rotate the orthogonal axes, in which a given components 
of EMT is evaluated under fixed $\k$ to make them coinciding with the 
direction of vectors $\U^r(t,\k)$.

In formulas (27) for spectral components of EMT the following notation
is introduced:
\beq
X^r=2S^r-\left(2S^r+U^r\right) {\cos^2(\theta)+1\over2}, \qquad
Y^r=-r\cos(\theta)V^r.
\eeq

The introducing of the functions $S^r$, $U^r$ and $V^r$ is carried out
in two steps. First we introduce the functions $\Phi^r$ and $\Psi^r$
according to formulas

\beq
\Y^r=\left( {kb\over\mu}\right)^{1/2} \left(\Phi^re_+ +\Psi^r
e_-\right), \quad \dot{\Y}^r=iK_0\left({kb\over\mu}\right)^{1/2}
\left(\Phi^re_+ -\Psi^re_-\right).
\eeq

The functions $\Phi^r$ and $\Psi^r$ satisfy the set of equations

\bea
\dot{\Psi}^r  &\!\!\!=\!\!\!& \Phi^r e_+^2 \left({W\over2} +ir
{\overline{W}\over2} \right)-ir{\overline{W}\over2}\Psi^r,\nonumber \\
\dot{\Phi}^r &\!\!\!=\!\!\!&  \Psi^re_-^2 \left({W\over2}
+ir{\overline{W}\over2} \right) +ir{\overline{W}\over2}\Phi^r,
\eea
where $W$ and $\overline{W}$ are defined by the relations

\beq
W=\dot{\mu}/\mu-\dot{b}/b, \quad \overline{W}= \overline{\Delta H} / \mu, \quad H_i\equiv
\A_i^{-1}\dot{\A}_1, \quad \overline{\Delta H} = H_3- H_1.
\eeq

Let us define the functions $S^r$, $U^r$ and $V^r$ as

\beq
S^r=|\Psi^r|^2, \quad U^r=2{\rm
Re}\left(\stackrel{\!\!\!\!\!\!*}{\Psi^r}\Phi^re_+^2\right), \quad V^r=2{\rm
Im}\left(\stackrel{\!\!\!\!\!\!*}{\Psi^r}\Phi^re_+^2\right).
\eeq
Differentiating these expressions by $t$ and taking into account (28),
we obtain a set of equations, satisfied by the functions
$S^r$, $U^r$, $V^r$:

\bea
\dot{S}^r &\!\!\!=\!\!\!& {W\over2}U^r +r{\overline{W}\over2}V^r,
\nonumber \\
\dot{U}^r &\!\!\!=\!\!\!& W(2S^r+1)-(r \overline{W}+2K_0)V^r,\nonumber \\
\dot{V}^r &\!\!\!=\!\!\!& r \overline{W}(2S^r+1) +(r \overline{W}+2K_0)U^r
\eea
with initial conditions $S^r=U^r=V^r=0$ at $t=t_0$. Moreover, there is
connection formula
\beq
(U^r)^2+(V^r)^2 =4S^r(S^r+1).
\eeq

Comparing the results for all types of fields, we get the conclusion
that the set of equations for functions $S^r$, $U^r$, $V^r$ can be
written in general form as

\be
{d\over dt}
\left( \ba{c}
S^r\\
U^r\\
V^r\\ \ea \right) =
\left( \ba{ccc}
0& {\ds W\over\ds 2} & r{\ds \overline{W}\over\ds2}\\
2W & 0& -(r \overline{W}+2K_0)\\
2rW& rW+2K_0& 0\\ \ea \right)
\left( \ba{c}
S^r\\
U^r\\
V^r\\ \ea \right) +
\left( \ba{c}
0\\
W\\
r \overline{W}\\ \ea \right).
\ee

As it can be seen, the structure of the set of equations (30) is similar
to that of the set of equations derived for scalar field and massive
spinor field [1,2].

The author is grateful to Yu.A.~Danilov, A.V.~Nesteruk, and
S.A.~Pritomanov for helpful discussions.

\end{document}